\documentclass[11pt,preprint]{aastex}

\begin{document}

\title{A Useful Approximation for Computing \\
the Continuum Polarization of Be Stars}

\author{David McDavid\altaffilmark{1}}
\affil{Limber Observatory, 135 Star Run, PO Box 63599,
Pipe Creek TX 78063}
\email{mcdavid@limber.org}
\altaffiltext{1}{Astronomical Institute, University of Amsterdam,
Netherlands}

\begin{abstract}
This paper describes a practical model for the polarization of Be
stars which can be used to estimate roughly the physical parameters
for optically thin circumstellar envelopes from broadband \mbox{$U\!BV\!RI$}\/
photopolarimetry data.  Analysis of long-term variability in terms of
these parameters is a promising approach toward understanding the Be
phenomenon.  An interesting result from fitting the model to
observations of eight Be stars is that all of them may have
geometrically thin disks, with opening half-angles on the order of ten
degrees or less.  This contributes to the growing evidence that most
Be disks are geometrically thin.
\end{abstract}


\section{Motivation}

The idea that the H$\alpha$ emission lines of Be stars originate from
an equatorially flattened circumstellar envelope or disk has now been
directly verified by interferometric imaging (Quirrenbach et al.\ 
1997).  The related explanation of the linear polarization of Be stars
as due to scattering of the starlight by free electrons in the disk
was also confirmed, because the observed position angle of the
polarization was found to be perpendicular to the direction of
elongation of the disk as projected onto the plane of the sky.  It is
this apparent asymmetry of the scattering region that results in a net
polarization, by defeating the cancellation which would take place if
there were identical scattering subregions in adjacent quadrants of
the projected disk (see Fig.~1).

\begin{figure}
\centerline{\includegraphics[width=4.5in]{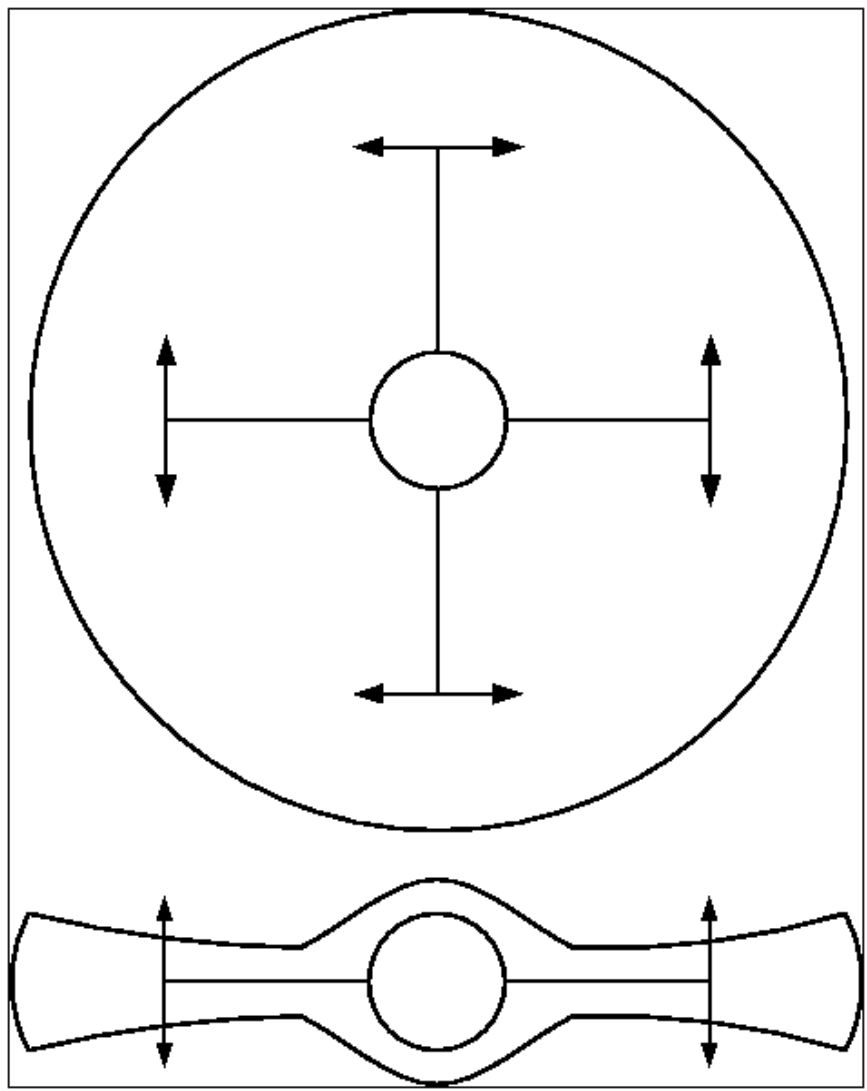}}
\caption{\footnotesize Scattering geometry in the disk of a Be star,
illustrating by extremes how the net polarization depends on the
orientation of the disk on the sky.  In the upper sketch the disk is
viewed pole-on, with angle of inclination to the line of sight
$i=0\arcdeg$, while in the lower sketch the disk is viewed equator-on,
with $i=90\arcdeg$.  Double arrows show the direction of vibration of
the electric field of starlight scattered in the disk.  For the
pole-on case cancellation occurs between adjacent quadrants of the
disk, and the net polarization is zero because the electric fields of
the scattered light are perpendicular.  When the disk is viewed
equator-on there is no cancellation, so the observed polarization is a
maximum.}
\end{figure}

A thorough historical review of both observation and theory of
polarization in Be stars was given by Coyne \& McLean (1982).  Since
that time the major observational developments in the field have been
increased spectral resolution by spectropolarimetry and extension of
the wavelength coverage into the ultraviolet by space-based
observations (Bjorkman 2000).  Beginning with the pioneering work of
Poeckert \& Marlborough (1978a,b), who calculated the single
scattering polarization in an NLTE hydrogen envelope, theoretical
analysis and modeling of the continuum polarization of Be star disks
has progressed through increasingly complicated treatments.  Fox
(1991) developed analytic equations to calculate the single scattering
polarization for axisymmetric envelopes. Bjorkman \& Bjorkman (1994)
extended the single scattering calculations to include attenuation and
emission within the envelope.  Hillier (1994) developed polarization
source functions using traditional Feutrier methods to integrate the
transfer equation for optically thick cases, while Wood, Bjorkman,
\& Bjorkman (1997, hereafter WBB) developed Monte Carlo methods.

One may wonder, then, if there is anything new to be learned about Be
stars from continued broadband polarimetry.  The answer is, of course,
that even though we now have firmer knowledge of the disk
characteristics, including temperature, density, and geometry,
broadband polarization monitoring is still a primary source of
information on the variability of these quantities over time scales of
years to decades, which are typical of the most fundamental aspect of
the Be phenomenon: the unpredictable transition from the normal B
phase to the Be phase and back again, associated with the formation
and later dissipation of the circumstellar envelope.

This paper attempts to establish a simple and approximate, but
nevertheless serviceable approach to the modeling and analysis of
\mbox{$U\!BV\!RI$}\/ polarimetry data to derive the basic envelope parameters.
Tracing the observed variability in terms of these parameters still
promises to give valuable hints about the nature of the unknown
processes involved in the Be phenomenon.

\section{A Spherical Sector Envelope Model}

Figure~2 shows the geometry of the circumstellar envelope model
adopted for this study.  Introduced by Kruszewski, Gehrels, \&
Serkowski (1968) to investigate the polarization of red variables, it
was later used by Brown \& McLean (1977) to illustrate their
theoretical formulation of the polarization by electron scattering in
a Be disk.  Its shape may be described as an axially symmetrical
sector of a sphere, with a wedge-shaped cross section opening outward
at half-angle $\alpha$.

\begin{figure}[htp]
\centerline{\includegraphics[width=3.5in]{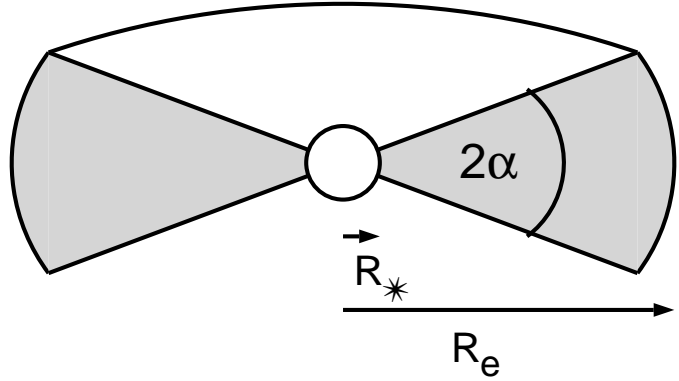}}
\caption{\footnotesize Cutaway view of the spherical sector model
adopted for the circumstellar envelope.}
\end{figure}

For small opening angles the spherical sector is a good representation
of an equatorial disk, with the advantage of being mathematically
suited for spherical coordinates.  A basic starting point is to assume
the disk is pure hydrogen with uniform electron temperature $T_{e}$,
extending to infinite distance ($R_{e}\rightarrow\infty$ in Fig.~2)
from a central star of radius $R_{*}$, with electron number density
$N_{e}$ given by a radial power law with exponent $\eta$\,:

\begin{equation}
N_{e}(r) = N_{0e} \left(\frac{R_{*}}{r}\right)^{\eta}\,.
\end{equation}

Waters, Cot\'{e}, \& Lamers (1987) derived values of $2.0<\eta<3.5$
for some of the stars also studied in this paper using the slope of
the infrared continuum from $IRAS$\/ observations, but the spherical
sector model with $R_{e}\rightarrow\infty$ has the peculiarity that
the disk mass is infinite unless $\eta>3.0$.  In what follows, the
symbol $\eta$ will be retained for generality, but $\eta=3.1$ is used
as a representative mean value in all the numerical computations.

\section{An Approximation for the Gray Polarization}

According to Fox (1991) except for a sign error in the first argument
of the beta function $B$, the net polarization of light from a central
star due to scattering of photons by free electrons (Thomson
scattering) in a surrounding envelope may be roughly approximated for
the specific case of the infinite spherical sector geometry as

\begin{equation}
p_{0}=\frac{3(\eta-1)}{16}B\left(\frac{\eta-1}{2},\frac{3}{2}\right)
\tau_{e}\sin\alpha\cos^{2}\alpha\sin^{2}i\,,
\end{equation}

\noindent where $i$ is the angle of inclination of the rotation axis
of the star to the line of sight and

\begin{eqnarray}
\tau_{e}&=&\int_{R_{*}}^{\infty}N_{0e}\left(\frac{R_{*}}{r}\right)^{\eta}
\sigma_{e} dr \nonumber \\
&=&\frac{N_{0e}\sigma_{e}R_{*}}{\eta - 1}
~\mbox{\rm (finite only for~} \eta>1.0)
\end{eqnarray}

\noindent is the total radial optical depth for electron scattering in
the equatorial plane of the spherical sector in terms of the Thomson
scattering cross section

\begin{equation}
\sigma_{e}=\frac{8\pi}{3}\left(\frac{e^2}{mc^2}\right)^2\,.
\end{equation}

Equation (2) is only a single-scattering approximation, based on the
assumption that the disk is optically thin and neglecting multiple
scattering.  It includes a correction for the geometrical complication
that the star is an extended source of light and cannot be treated as
a point source (Cassinelli, Nordsieck, \& Murison 1987, hereafter
CNM), but it does not take into account the occultation of part of the
scattering envelope by the star.

\section{Including the Wavelength Dependence}

So far the calculated polarization has no dependence on wavelength,
since it comes from pure electron scattering.  However, neutral
hydrogen in the envelope absorbs some light both before and after
scattering, and the envelope emission (which is treated here as
unpolarized and is not considered to scatter) dilutes the
polarization.  It is therefore necessary to include these
wavelength-dependent effects, since they combine to produce the slope
of the continuum polarization and the well-known abrupt changes in the
polarization at the \ion{H}{1} ionization series limits, as first
demonstrated by Capps, Coyne, \& Dyck (1973) and later refined by
McLean (1979).

Assuming LTE ionization fractions and level populations, equation~(5)
gives the total wavelength-dependent opacity $\kappa(\lambda)$ per
gram of neutral hydrogen in the ground state (Aller 1963).  $C_{0}$ is
a fixed numerical factor, $X(n)$ is the ionization energy from level
$n$ in units of $kT_{e}$, and $C_{se}$ is the correction factor for
stimulated emission. The three individual terms in the equation are a
summation over the ionization edges of the first seven discrete energy
levels, an integrated term for the combination of all the remaining
levels, and a free-free absorption term (with Gaunt factors taken to
be unity).  See Appendix~A for a discussion of NLTE corrections.

\begin{equation}
\kappa(\lambda)=C_{0}e^{-X(1)}
\lambda^{3}\left(\sum_{\lambda<\lambda_{n}}^{X(n)}
\frac{e^{X(n)}}{n^{3}}+
\frac{(e^{X(8)}-1)}{2X(1)}+
\frac{1}{2X(1)}
\right)C_{se}\,,
\end{equation}

\noindent where

\begin{equation}
C_{0}=\frac{32\pi^{2}e^{6}R}{3\sqrt{3}m_{H}h^{3}c^{3}}\,,
\end{equation}

\begin{equation}
X(n)=\frac{2\pi^{2}m_{e}e^{4}}{n^{2} h^{2} k T_{e}}\,,
\end{equation}

\noindent and

\begin{equation}
C_{se}=\left(1-e^{-hc/\lambda kT_{e}}\right)\,.
\end{equation}

If the free electron number density $N_{e}(r)$ and the electron
temperature $T_{e}$ are known, then the number density $N_{1}(r)$ of
neutral hydrogen atoms in the ground state can be found from the Saha
equation:

\begin{eqnarray}
N_{1}(r)&=&\frac{h^{3}}{(2\pi m_{e}kT_{e})^{3/2}}N_{e}^{2}(r)e^{X(1)}
\nonumber \\
&=&N_{01}\left(\frac{R_{*}}{r}\right)^{2\eta}\,,
\end{eqnarray}

\noindent where

\begin{equation}
N_{01}=\frac{h^{3}}{(2\pi m_{e}kT_{e})^{3/2}}N_{0e}^{2}e^{X(1)}\,.
\end{equation}

\noindent The total radial optical depth for neutral hydrogen absorption
in the equatorial plane of the spherical sector is therefore

\begin{eqnarray}
\tau_{a}(\lambda)&=&\int_{R_{*}}^{\infty}\kappa(\lambda)m_{H}
N_{01}\left(\frac{R_{*}}{r}\right)^{2\eta} dr
\nonumber \\
&=&\frac{N_{01}m_{H}\kappa(\lambda)R_{*}}{2\eta-1}
~\mbox{\rm (finite only for~} \eta>0.5)\,,
\end{eqnarray}

\noindent which acts as a wavelength-dependent attenuation factor
to reduce the gray polarization.

The volume emission coefficient of the envelope is

\begin{equation}
j(\lambda,r)=m_{H}\kappa(\lambda)
N_{01}\left({\frac{R_{*}}{r}}\right)^{2\eta}B(\lambda)\,,
\end{equation}

\noindent where

\begin{equation}
B(\lambda)=\frac{2hc^{2}}{\lambda^{5}(e^{hc/\lambda kT_{e}}-1)}
\end{equation}

\noindent is the Planck function (not to be confused with the beta
function in \S3).  The total luminosity of the envelope is then

\begin{eqnarray}
L(\lambda)&=&4\pi\int_{0}^{2\pi}\int_{\frac{\pi}{2}-\alpha}^{\frac{\pi}{2}+
\alpha}\int_{R_{*}}^{\infty}j(\lambda,r)r^2\sin\theta dr d\theta d\phi
\nonumber \\
&=&\frac{16\pi^{2}}{2\eta-3}N_{01}m_{H}\kappa(\lambda)
B(\lambda)R_{*}^{3}\sin\alpha ~\mbox{\rm (finite only for~} \eta>1.5)\,,
\end{eqnarray}

\noindent which together with the stellar flux $F_{*}(\lambda)$ produces
a further wavelength dependence of the polarization.  Theoretical values
of the stellar flux $F_{*}(\lambda)$ are taken from tabulated model
atmospheres by Kurucz (1994).

With the wavelength dependence included, our polarization estimate may
now be written as

\begin{equation}
p(\lambda) = \frac{p_{0}e^{-\tau_{a}(\lambda)}}
{1+L(\lambda)/(4\pi R_{*}^{2}F_{*}(\lambda))}\,.
\end{equation}

It should be emphasized that this equation is only an approximation
(even if the disk is optically thin), since it includes neither the
contribution of scattering of the disk emission to the polarization
nor attenuation of the direct starlight by the disk.  Also the neutral
hydrogen opacity is treated in a very crude way, using only the
maximum radial optical depth in the equatorial plane.

It may be of interest to calculate the total mass of the disk, which
can be done approximately by simply counting pairs of electrons and
protons under the assumption that it consists purely of fully ionized
hydrogen:

\begin{eqnarray}
M&=&\int_{0}^{2\pi}\int_{\frac{\pi}{2}-\alpha}^{\frac{\pi}{2}+\alpha}
\int_{R_{*}}^{\infty}m_{H}N_{0e}\left(\frac{R_{*}}{r}\right)
^{\eta}r^2\sin\theta dr d\theta d\phi \nonumber \\
&=&\frac{4\pi}{\eta-3}N_{0e}m_{H}R_{*}^{3}\sin\alpha
~\mbox{\rm (finite only for~} \eta>3.0)\,.
\end{eqnarray}

\section{Synthesizing Broadband Data}

Adjustment of the envelope parameters in the theoretical model to give
results consistent with observations might be expected to yield
valuable information about the nature of Be disks.  To compare with
observations on the Johnson-Cousins \mbox{$U\!BV\!RI$}\/ system, it is
first necessary to convolve the theoretically estimated polarization
p($\lambda$) from equation~(15) with the bandpass characteristics of
the optical system.

As a first approximation, let us assume Gaussian filter transmission
curves based on the standard \mbox{$U\!BV\!RI$}\/ wavelengths and
$fwhm$ (Bessell 1979), which are closely matched by the system
actually used for the observations (see Table~1).

\begin{deluxetable}{ccc}
\tabletypesize{\small}
\tablecolumns{3}
\tablewidth{0pt}
\tablecaption{Filter System Parameters}
\tablehead{
\colhead{Filter} &
\colhead{Effective Wavelength} &
\colhead{Bandpass ($fwhm$)} \\
& (\AA) & (\AA)
}

\startdata

$U$ & 3650 & \phn700 \\
$B$ & 4400 & 1000 \\
$V$ & 5500 & \phn900 \\
$R$ & 6400 & 1500 \\
$I$ & 7900 & 1500 \\

\enddata
\end{deluxetable}

Given the central wavelength $\lambda_{c}$, the standard deviation
$\sigma$ can be calculated from the $fwhm~\Gamma$ since $\Gamma =
2.354~\sigma$ for a Gaussian distribution, so that the transmission
function may be written as

\begin{equation}
T(\lambda)=\exp\left[-\frac{1}{2}\left(\frac{\lambda-\lambda_{c}}
{\Gamma/2.354}\right)^2\right]\,.
\end{equation}

The theoretical total flux $F(\lambda)$ is due to the stellar
component $F_{*}(\lambda)$ plus the envelope component:

\begin{equation}
F(\lambda)=F_{*}(\lambda)+L(\lambda)/(4 \pi R_{*}^{2})\,.
\end{equation}

Convolving the polarization in terms of flux with the transmission
function, where $\lambda_{min}$ and $\lambda_{max}$ are, to a
realistic approximation, the 10\% response points, we find a
theoretical expression for the expected polarization over the given
bandpass:

\begin{equation}
P(BP)=\frac{\sum_{\lambda=\lambda_{min}}^{\lambda_{max}}p(\lambda)
F(\lambda)T(\lambda)\Delta\lambda}{\sum_{\lambda=\lambda_{min}}^
{\lambda_{max}}F(\lambda)T(\lambda)\Delta\lambda}\,.
\end{equation}

\section{Adjustable Parameters \& Model Fits}

With the model in hand, an interactive graphical computer interface
was designed to allow adjusting the input parameters by trial and
error to fit \mbox{$U\!BV\!RI$}\/ polarization measurements (McDavid
1999) of eight Be stars.  Solutions obtained this way should be viewed
with some caution since they may not be unique: different geometrical
distributions of scatterers can produce the same net polarization.
Moreover, given the gross approximations involved, only
order-of-magnitude results should be expected.

The model disk has three adjustable parameters: the maximum electron
number density $N_{0e}$, the angle of inclination $i$ of the rotation
axis to the line of sight, and the opening half-angle $\alpha$.
Additionally required fixed parameters for each individual star are
the radius $R_{*}$, the effective temperature $T_{*}$, and the surface
gravity $\log~g$, which were estimated from Table~1 of Collins, Truax,
\& Cranmer (1991) based on spectral types from Slettebak (1982).  The
electron temperature of the disk was then fixed at $T_{e}=0.75~T_{*}$
and the appropriate Kurucz flux table chosen to match $T_{*}$ and
$\log~g$.

The fixed disk temperature $T_{e}$ affects the wavelength dependence
of the polarization through its influence on the neutral hydrogen
opacity by setting the degree of ionization and the populations of the
excited states (eq.~[5]).  It influences not only the hydrogen
absorption optical depth (eq.~[11]), but also the disk luminosity
(eq.~[14]), which is another contributor to the wavelength dependence
of the polarization.  As a result, $T_{e}$ is important in determining
the polarization Balmer jump and the slope of the polarization over
the Paschen continuum, both of which generally increase at lower
temperatures.  Appendix~A explains the NLTE corrections to the level
populations which are necessary to fit these features for some of the
program stars.

It is instructive to summarize individually the effects of the three
adjustable parameters on the behavior of $p(\lambda)$ according to
equation~(15).

(1) The overall degree of gray polarization is directly proportional
to $N_{e0}$ through the factor $\tau_{e}$ in equation~(2).  However,
simply increasing $N_{e0}$ does not always result in an overall
polarization increase, because it also increases the attenuation and
adds to the wavelength dependence of the polarization through
$\tau_{a}(\lambda)$ (eq.~[11]) and $L(\lambda)$ (eq.~[14]).

(2) The earliest basic electron scattering models (e.g.~Brown \&
McLean 1977) gave a $\sin^{2}i$ dependence of the polarization on $i$,
and further refinements have not qualitatively changed the simple
picture of Figure~1 in which a face-on disk shows no net polarization
because of symmetrical cancellation, while the maximum asymmetry of an
edge-on disk results in the maximum polarization.  It may seem likely
that $i$ is poorly determined in this model since the same
polarization might be obtained, for example, by decreasing $i$ while
increasing $\alpha$, keeping $N_{e0}$ constant.  Experiments do not
bear this out, though, because a thicker disk has a higher luminosity,
which changes the wavelength dependence of the polarization enough to
ruin the fit (eq.~[15]).  Another helpful constraint on $i$ is the
presence or absence of shell lines, which are thought to be absorption
features in the spectral line profiles due to circumstellar material
in the line of sight for nearly edge-on equatorial disks.  Since Be
stars are generally rapid rotators, spectroscopic measurement of the
projected rotational velocity $v\,\sin\,i$ can also suggest roughly
the value of~$i$.

(3) Figure~3 shows how the model polarization $p^{B}$ in the $B$
passband varies as $\alpha$ increases from $0\arcdeg$ to $90\arcdeg$
using a typical set of parameters.  The polarization first rises as
the widening disk presents an increasing number of scatterers, then
begins to decline due to the buildup of enough density at high
latitudes to cancel the polarization from the equatorial regions.  The
model of Waters \& Marlborough (1992) also shows this behavior, which
becomes of practical concern because the hump-shaped graph allows for
both a ``thick disk'' and a ``thin disk'' solution, producing
identical polarization at $\alpha$ values on opposite sides of the
peak.  For equal densities, however, the thick disk will have a higher
luminosity than the thin disk, giving the polarization a recognizably
different wavelength dependence.

\begin{figure}[htp]
\centerline{\includegraphics[width=3.5in]{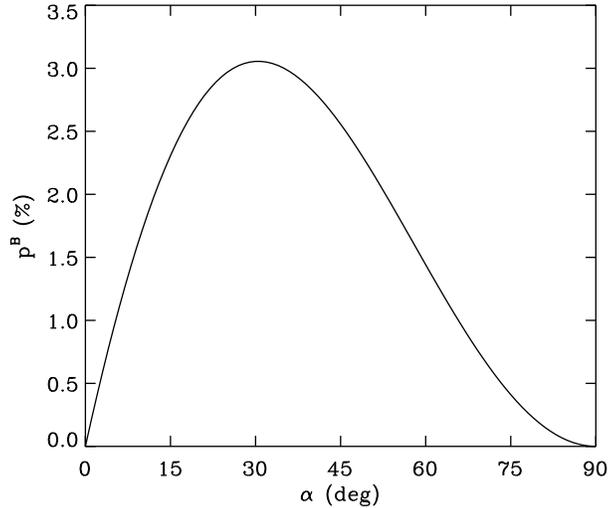}}
\caption{\footnotesize The dependence of the model blue polarization
$p^{B}$ on the opening half-angle $\alpha$ of the disk using a
typical set of parameters.}
\end{figure}

Figures~4--11 present possible (but not necessarily unique) model fits
for the eight program Be stars, with the parameters summarized in
Table~2.  The observational data points, plotted as open squares with
vertical error bars, are intrinsic polarization, corrected for the
interstellar component by Stokes vector subtraction (McDavid 1999).
The corresponding model result is plotted as a solid line, using open
circles for the broadband values with horizontal line segments to show
the filter passbands.  Several additional quantities derived from the
model are also included in Table~2 for each fit: the maximum
polarization $pmax$ and its wavelength $\lambda(pmax)$, the maximum of
the \ion{H}{1} absorption optical depth $\tau_{a}max$ and its
wavelength $\lambda(\tau_{a}max)$, the electron scattering optical
depth $\tau_{e}$, the total envelope mass $M$, and the NLTE departure
coefficients $b_{2}$ and $b_{3}$ for the populations of the first two
excited states of the neutral hydrogen in the disk.  Refer to
Appendix~A for a discussion of NLTE considerations.

\begin{figure}
\centerline{\includegraphics[width=3.5in]{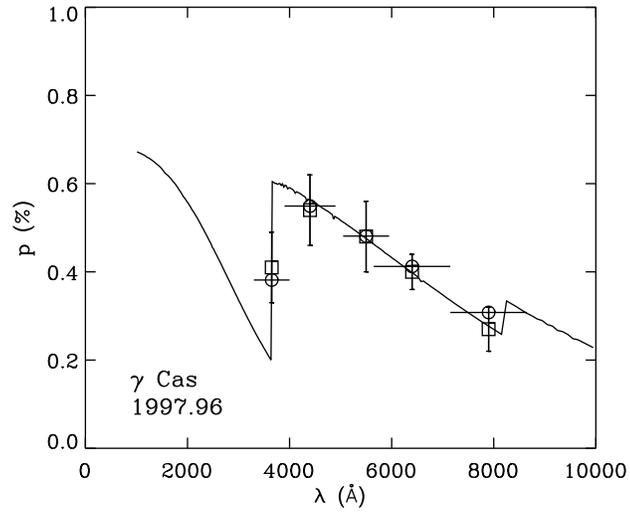}}
\caption{\footnotesize  Fit of a spherical sector disk
to the \mbox{$U\!BV\!RI$}\/ polarization of $\gamma$~Cas.}
\end{figure}

\begin{figure}
\centerline{\includegraphics[width=3.5in]{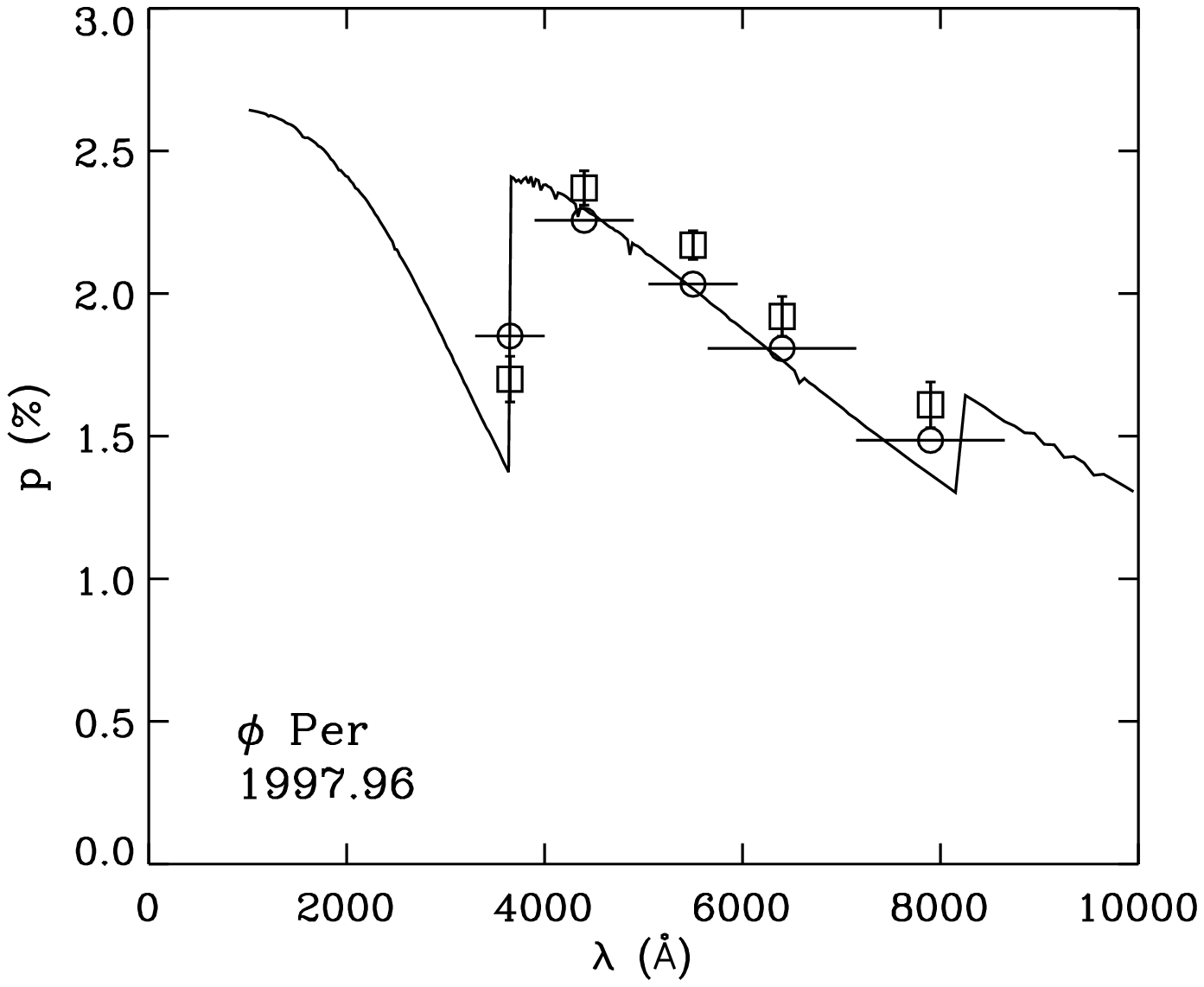}}
\caption{\footnotesize  Fit of a spherical sector disk
to the \mbox{$U\!BV\!RI$}\/ polarization of $\phi$~Per.}
\end{figure}

\begin{figure}
\centerline{\includegraphics[width=3.5in]{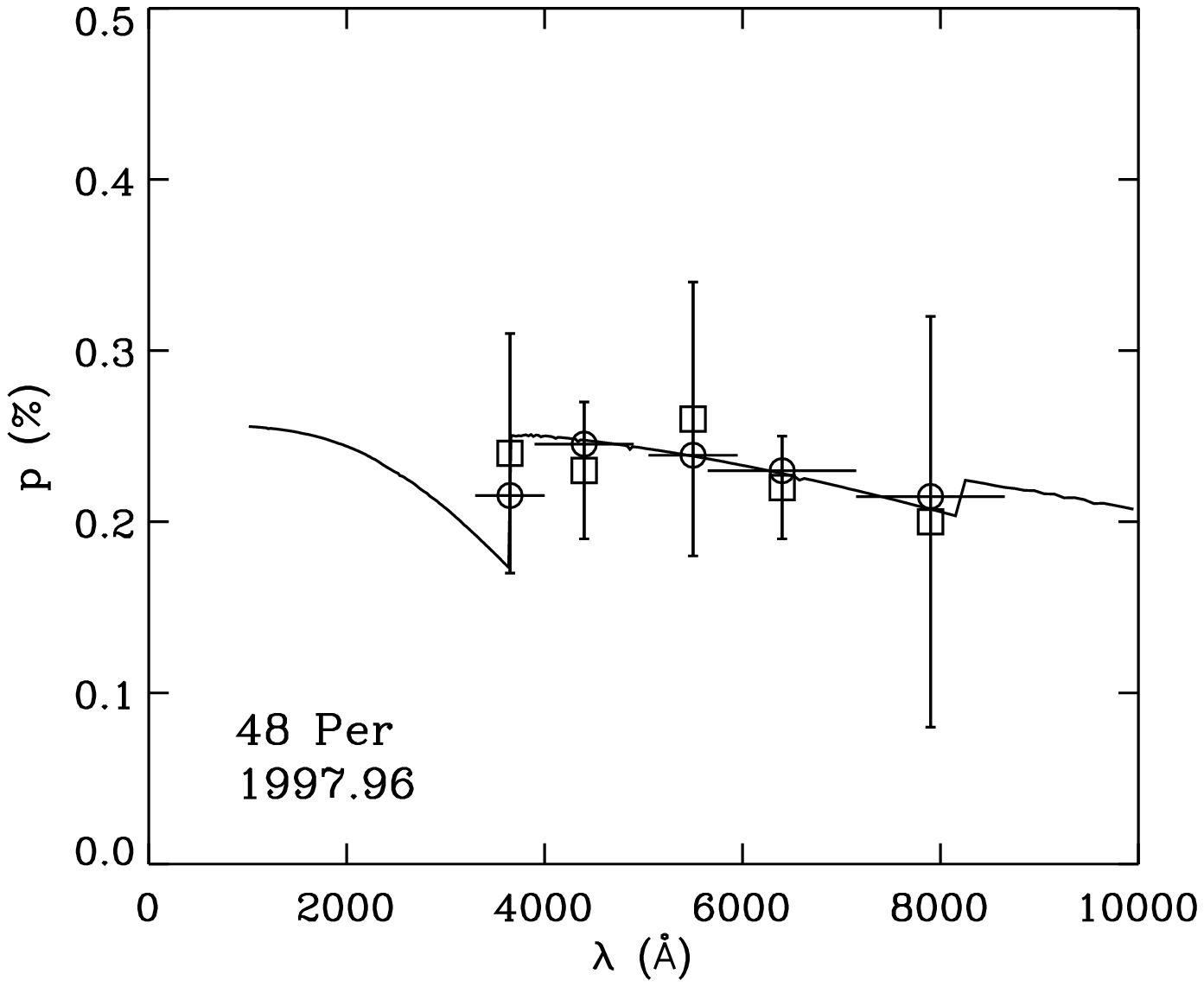}}
\caption{\footnotesize  Fit of a spherical sector disk
to the \mbox{$U\!BV\!RI$}\/ polarization of 48~Per.}
\end{figure}

\begin{figure}
\centerline{\includegraphics[width=3.5in]{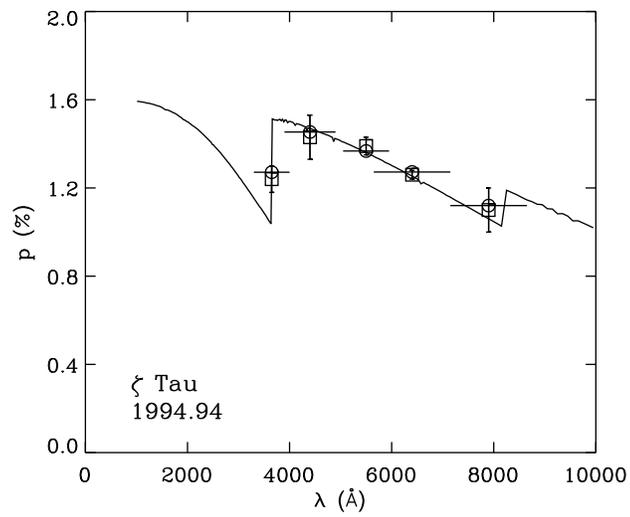}}
\caption{\footnotesize  Fit of a spherical sector disk
to the \mbox{$U\!BV\!RI$}\/ polarization of $\zeta$~Tau.}
\end{figure}

\begin{figure}
\centerline{\includegraphics[width=3.5in]{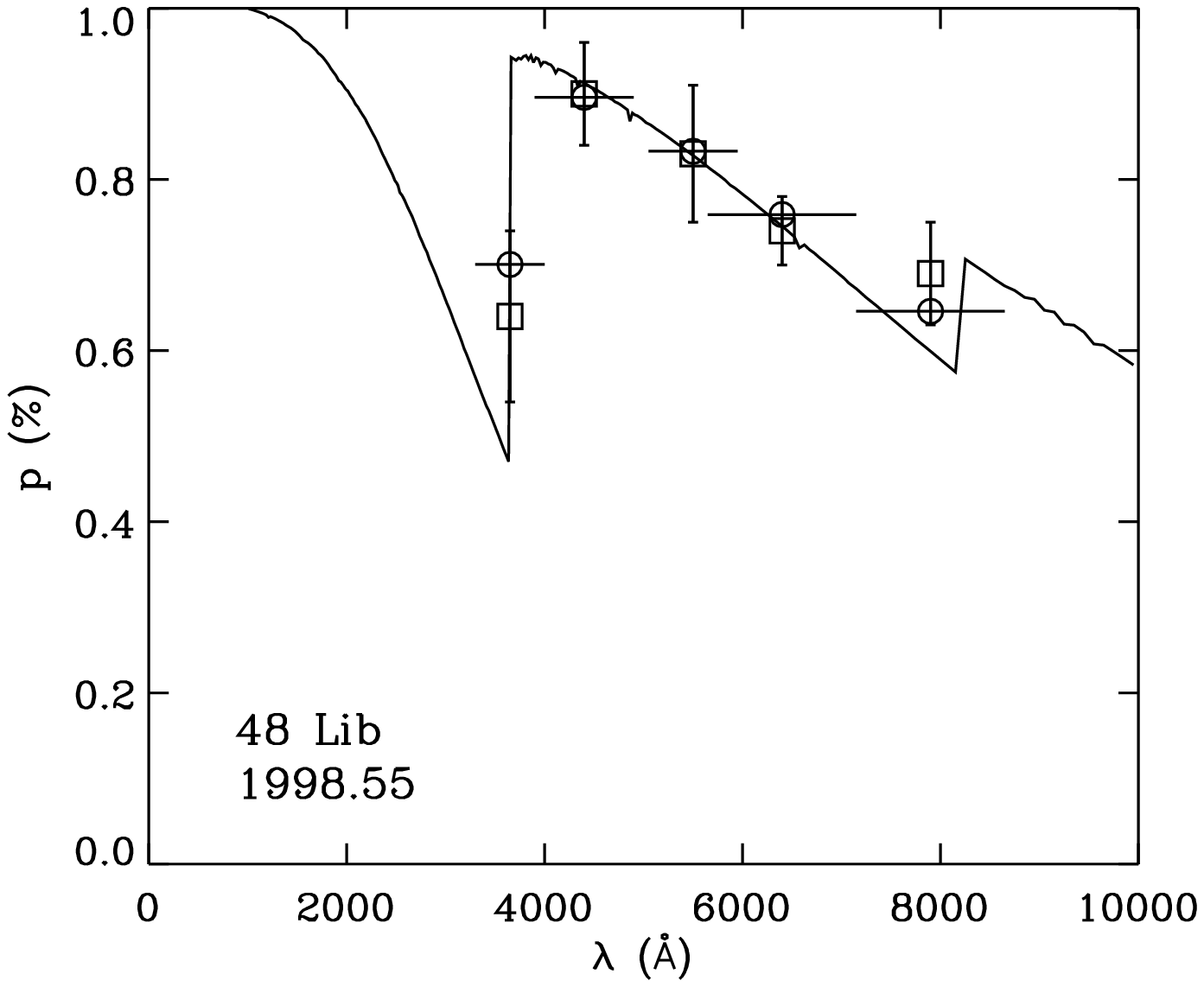}}
\caption{\footnotesize  Fit of a spherical sector disk
to the \mbox{$U\!BV\!RI$}\/ polarization of 48~Lib.}
\end{figure}

\begin{figure}
\centerline{\includegraphics[width=3.5in]{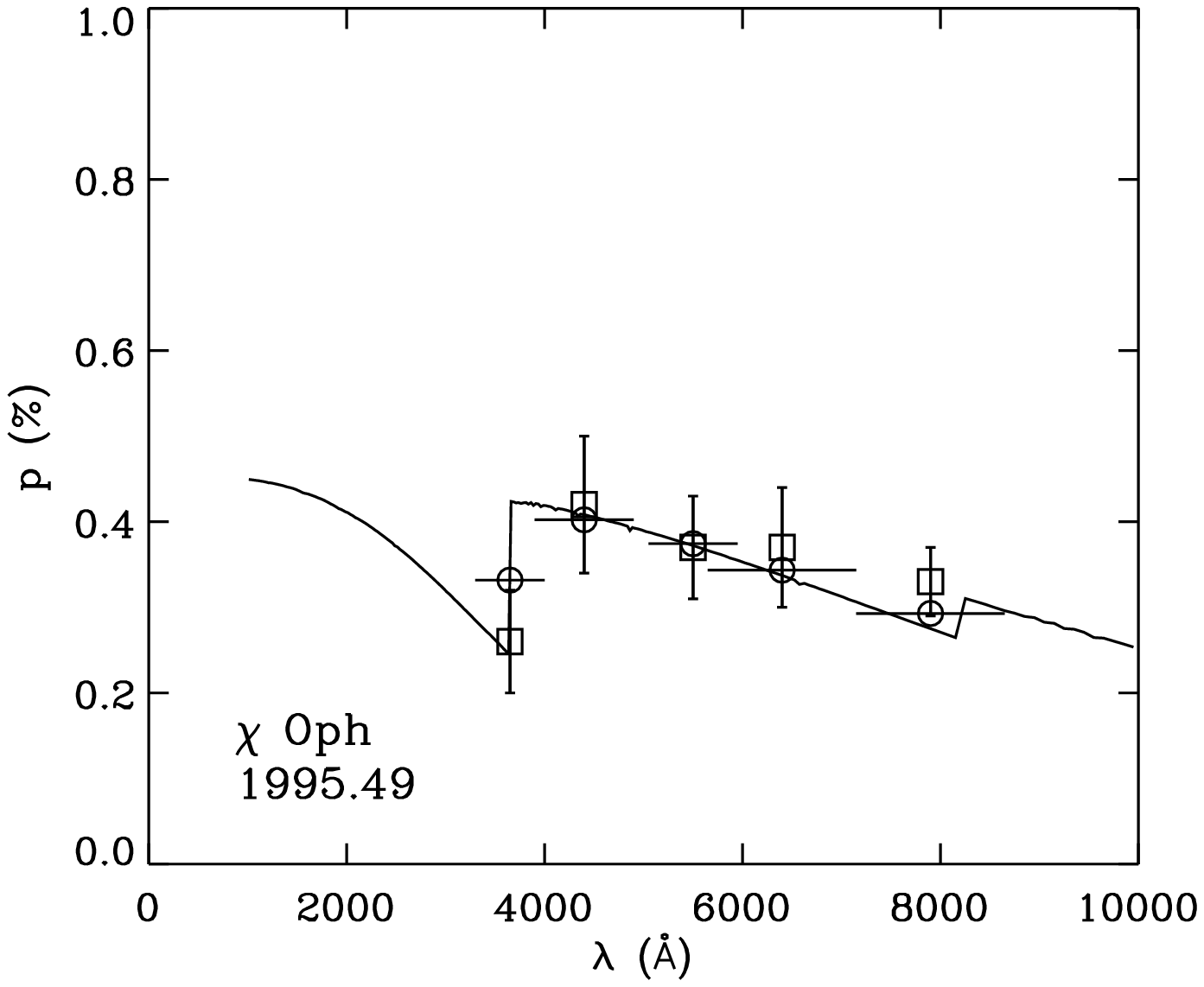}}
\caption{\footnotesize  Fit of a spherical sector disk
to the \mbox{$U\!BV\!RI$}\/ polarization of $\chi$~Oph.}
\end{figure}

\begin{figure}
\centerline{\includegraphics[width=3.5in]{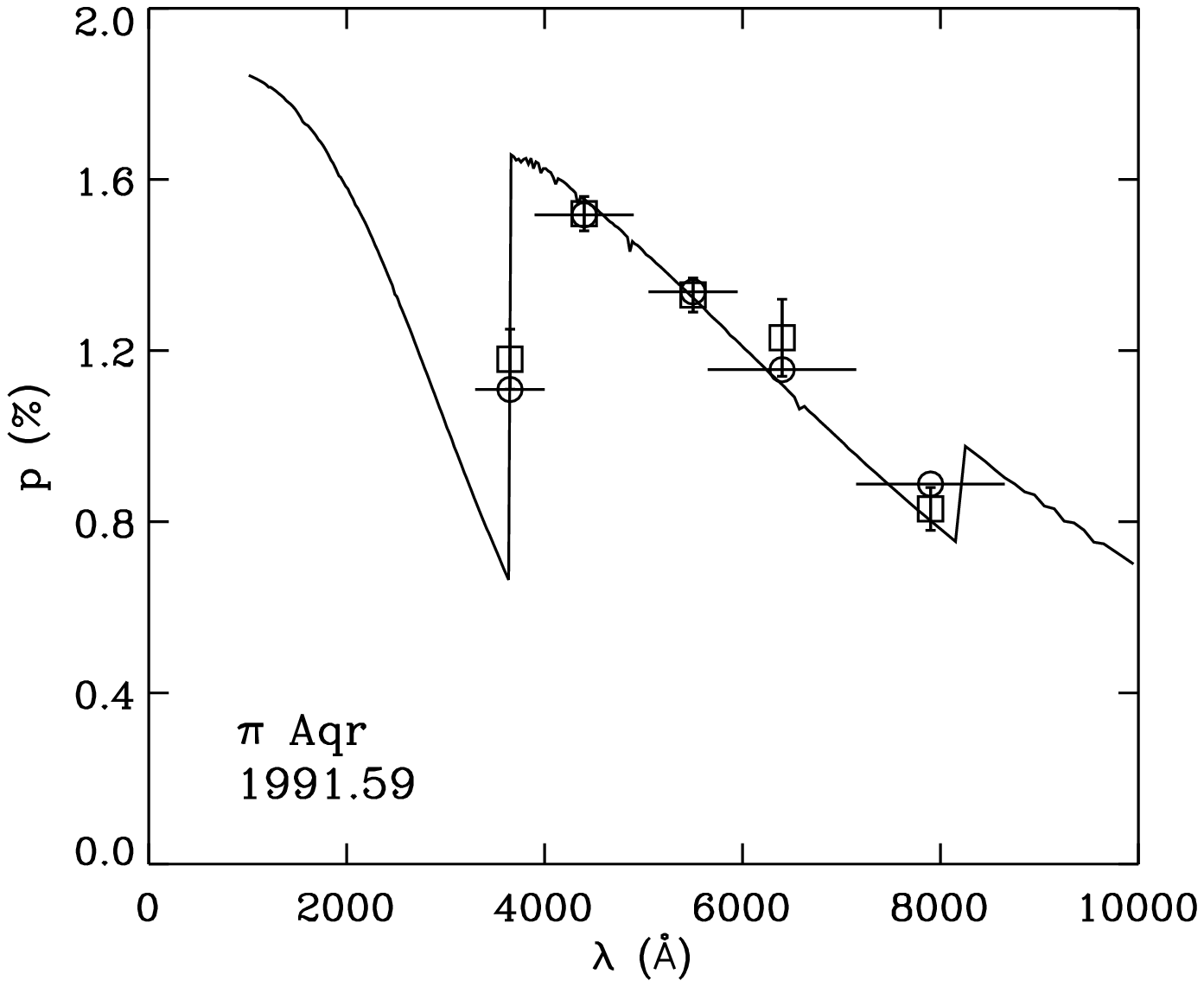}}
\caption{\footnotesize  Fit of a spherical sector disk
to the \mbox{$U\!BV\!RI$}\/ polarization of $\pi$~Aqr.}
\end{figure}

\begin{figure}
\centerline{\includegraphics[width=3.5in]{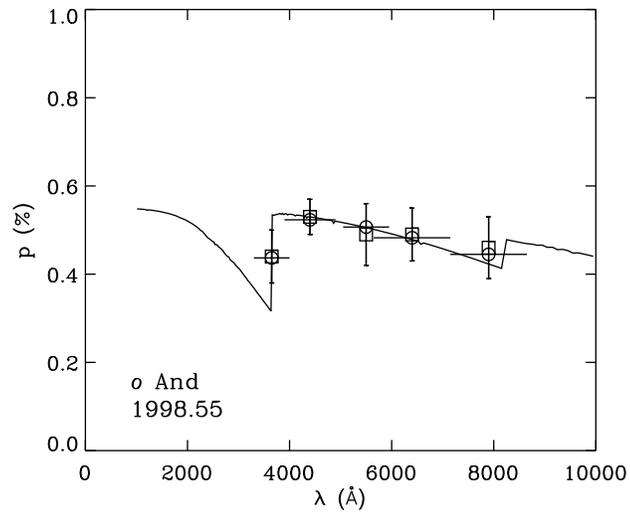}}
\caption{\footnotesize  Fit of a spherical sector disk
to the \mbox{$U\!BV\!RI$}\/ polarization of o~And.}
\end{figure}

\begin{deluxetable}{lcccccccc}
\tabletypesize{\small}
\rotate
\tablewidth{0pt}
\tablecolumns{9}
\tablecaption{Be Star Modeling Data}
\tablehead{
\colhead{} &
\colhead{$\gamma$ Cas} &
\colhead{$\phi$ Per} &
\colhead{48 Per} &
\colhead{$\zeta$ Tau} &
\colhead{48 Lib} &
\colhead{$\chi$ Oph} &
\colhead{$\pi$ Aqr} &
\colhead{o And}
}

\startdata

Spectral Type & B0.5 IVe & B1.5 (V:)e-shell & B4 Ve & B1 IVe-shell &
B3:IV:e-shell & B1.5 Ve & B1 III-IVe & B6 III \\

Date of Obs. & 1997.96 & 1997.96 & 1997.96 & 1994.94 & 1998.55 &
1995.49 & 1991.59 & 1998.55 \\

$R_{*} (R_{\sun})$ & 6.5 & 6.0 & 4.5 & 7.0 & 6.0 & 6.0 & 9.0 & 5.5 \\

$T_{*}$ (K) & 25,000 & 25,000 & 17,000 & 25,000 & 20,000 & 25,000 &
25,000 & 14,000 \\

$\log g$ & 4.0 & 4.0 & 4.0 & 4.0 & 3.5 & 4.0 & 3.5 & 3.5 \\

$N_{0e} (10^{12}{\rm cm}^{-3})$ & 9.00 & 5.00 & 3.00 & 4.20 & 4.60 & 6.20 &
6.50 & 2.00 \\

$T_{e}$ (K) & 18,750 & 18,750 & 12,750 & 18,750 & 15,000 & 18,750 &
18,750 & 10,500 \\

$\alpha$(deg) & 2.0 & 10.0 & 3.0 & 6.0 & 4.0 & 3.0 & 3.5 & 5.5 \\

$i$(deg) & 52.0 & 80.0 & 55.0 & 80.0 & 80.0 & 41.0 & 80.0 & 80.0 \\

$pmax$(\%) & 0.60 & 2.41 & 0.25 & 1.51 & 0.94 & 0.42 & 1.66 & 0.54 \\

$\lambda(pmax)$ (\AA) & 3661 & 3862 & 3862 & 3661 & 3812 & 3661 &
3661 & 3862 \\

$\tau_{a}max$ & 0.94 & 0.27 & 0.23 & 0.22 & 0.44 & 0.41 & 0.68 & 0.22 \\

$\lambda(\tau_{a}max)$ (\AA) & 3636 & 3636 & 3636 & 3636 & 3636 & 3636 &
3636 & 3636 \\

$\tau_{e}$ & 1.29 & 0.66 & 0.30 & 0.65 & 0.61 & 0.82 & 1.29 & 0.24 \\

$M (10^{-9}M_{\sun})$ & 3.07 & 6.69 & 0.51 & 5.37 & 2.47 & 2.50 &
10.3 & 1.14 \\

$b_{2}$ & 1.69 & 1.69 & 1.05 & 1.69 & 1.35 & 1.69 & 1.69 & 0.70 \\

$b_{3}$ & 0.29 & 0.29 & 0.23 & 0.29 & 0.26 & 0.29 & 0.29 & 0.19 \\

\enddata
\end{deluxetable}

\section{Test of Accuracy}

Since the model presented here is such a crude approximation, it is
important to compare its results with those from a more sophisticated
model to investigate, at least qualitatively, the accuracy we can
expect.  The Monte Carlo code of WBB is one such standard, and it is
based on the same set of input parameters except that the density
input is the electron scattering optical depth rather than the
electron number density, which requires only a minor conversion.
Ideally we would like to determine WBB fits to the program star
observations, and then quantitatively evaluate the simplified model by
direct comparison of the fit parameters.  Unfortunately, because of
the slow convergence of the Monte Carlo method, deriving a WBB fit is
a complex process requiring substantial amounts of cpu time even on a
fast computer.  Otherwise there would be no reason for interest in the
kind of quick approximation that is the subject of this paper.

For the time being, though, we can at least make use of the existing
WBB fit to $\zeta$~Tau.  Figure~12 shows examples of $p(\lambda)$ for
the quick approximation (solid line with broadband filter points
marked by open circles) compared to the full optically thick multiple
scattering WBB simulation (dashed line connecting filled circles).
Errors in the WBB data points are on the order of 0.03\%, limited by
computing time.  The observed \mbox{$U\!BV\!RI$}\/ data are plotted as
open squares with error bars.  Using the parameters
$R_{*}=5.5~R_{\sun}$, $T_{*}=19,000$~K, $T_{e}=15,000$~K,
$\alpha=2\fdg5$, and $i=82\fdg0$ from the WBB solution as input to
both models, the six plots show what happens as $\tau_{e}$ is
increased by steps from 0.10 to 3.00 (the best fit value according to
WBB).

\begin{figure}
\centerline{\includegraphics[width=6.0in]{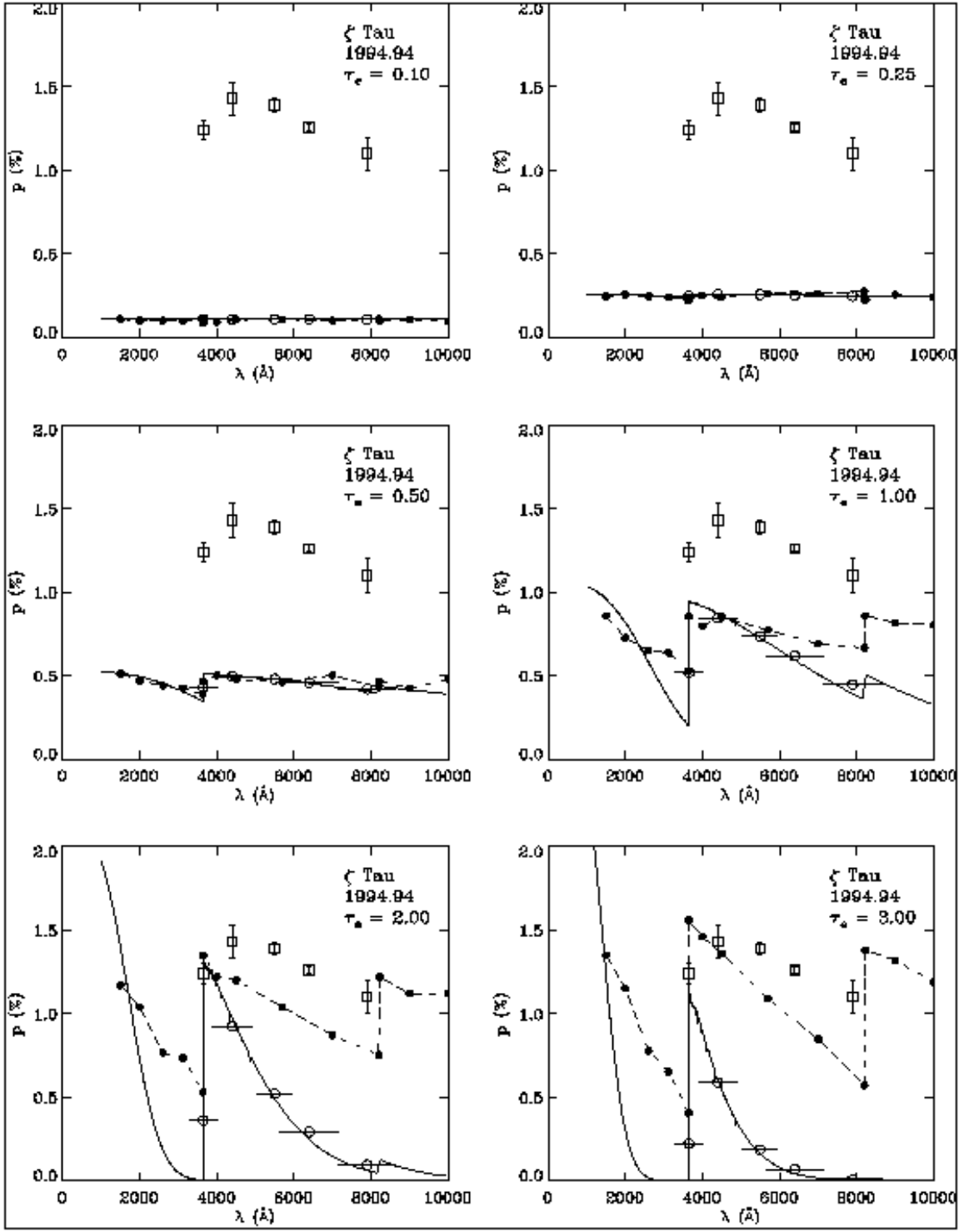}}
\caption{\footnotesize  A comparison of the simple approximation
(solid line with open circles) against the WBB solution (dashed line
with filled circles) for a range of electron scattering optical depths
$\tau_{e}$.  The observational data points for $\zeta$~Tau are shown
as open squares.}
\end{figure}

This test shows that the optically thin approximation agrees quite
well with the WBB calculation up to and including an optical depth of
1.00.  It is also encouraging that for $\zeta$~Tau the best fit of the
approximate method as shown in Figure~7 and Table~2 has
$N_{e0}=4.20 \times 10^{12}~{\rm cm^{-3}}$, which is well within the
range of $2.38 \times 10^{12} \leq N_{0e} \leq 7.54 \times 10^{12}~{\rm
cm^{-3}}$ found by Waters, Cot\'{e}, \& Lamers (1987) in their
infrared study.

However, it must be noted that there is a possibly serious
disagreement between the WBB model and the simplified model presented
here with regard to the best-fit values of $\tau_{e}$ and $\alpha$ for
$\zeta$~Tau.  While the WBB disk ($\tau_{e}=3.00$ and $\alpha=2\fdg5$)
is optically thick but geometrically thin, the simplified model disk
($\tau_{e}=0.65$ and $\alpha=6\fdg0$) is optically thin and
geometrically thicker (although it still qualifies as ``geometrically
thin'').

In Figure~12 the approximate model fails to produce a polarization
large enough to fit the observations of $\zeta$~Tau by increasing
$\tau_{e}$ at constant $\alpha$ because it becomes dominated by
\ion{H}{1} opacity.  As shown in Table~2, this problem is seemingly
avoided by increasing $\alpha$ instead of $\tau_{e}$, but the
resulting optically thin solution could possibly be ruled out on other
grounds, such as the associated infrared emission as discussed by WBB\@.
This is only one cautionary example emphasizing the importance of
independent observational consistency checks in the interpretation of
Be-star polarimetry data.

\section{Conclusion}

In summary, the model presented here appears to give reasonable
results for optically thin cases, but we should only trust it for
order-of-magnitude accuracy because it is based on so many simplifying
assumptions.  Even so, it may be useful for estimating trial values of
the parameters as starting points for rigorous fitting with more
elaborate models.

An interesting result from applying the model to observations of the
eight program stars is that all of them may be fit with geometrically
thin disks, with opening half-angles of ten degrees or less.  This
adds tentative support to the statistics in favor of thin disks as
presented by Bjorkman \& Cassinelli (1990) and later discussed at
length by WBB\@.  If most Be disks prove to be similarly flat, it will
surely have important implications concerning their formation process.

It may seem discouraging that this model will never find optically
thick solutions even if they physically exist.  And with three
adjustable parameters, there might be any number of apparently valid
optically thin fits to a set of \mbox{$U\!BV\!RI$}\/ data points,
rendering the entire exercise inconclusive.  Nevertheless, there are
good reasons for a more optimistic point of view.  Due to the efforts
of many investigators, as documented, for example, in the references
in \S1 of this paper, the range of physically allowable parameters has
been narrowed significantly.  And within even the most liberal of
these constraints, experience with a quick, interactive
trial-and-error system based on even the elementary physics presented
here will show that it is surprisingly difficult to find multiple
solutions with qualitatively different sets of parameters.  I will be
glad to share my IDL modeling program with interested users on
request.

\acknowledgments

Thanks to Joe Cassinelli for critical reading and commentary on early
drafts of this paper, and especially for suggesting a method to
estimate the NLTE departure coefficients for the neutral hydrogen in a
Be disk.  Kenny Wood and Barbara Whitney also provided very helpful
discussions, and Jon Bjorkman kindly gave me a version of the Monte
Carlo polarization code.  I especially thank the anonymous referee for
pointing out many serious errors in the original manuscript and for
patiently explaining the necessary corrections.

\appendix

\section{Estimating the NLTE Departure Coefficients}

Fits to the polarization Balmer jump were difficult and in some cases
impossible without NLTE corrections to the level populations of the
first two excited states of \ion{H}{1}.\  According to the combined
Saha and Boltzmann equations, the maximum population of the first
excited state normally occurs at a temperature of about 10,000~K and
declines rapidly toward higher temperatures with increasing excitation
and ionization.  At temperatures closer to 20,000~K appropriate for
Be disks, the LTE value of the absorption coefficient at the $U$-band
is sometimes too small to match the large polarization Balmer jumps
which are commonly observed.  NLTE correction is applied only to the
first two excited states because they dominate the opacity at optical
wavelengths for the conditions of density and temperature of interest.

Since only averages over a line of sight are required for this very
approximate polarization model, departure coefficients were simply
estimated from the calculations of CNM\@.  Their Figure~3 shows mean
population parameters $\overline{q}_{n}$ for $n=2$ and $n=3$ defined
by

\begin{eqnarray}
\overline{q}_{nNLTE}&\equiv&\frac{\int_{R_{*}}^{\infty}
q_{n}(r,T_{*})N_{e}^{2}dr}{\int_{R_{*}}^{\infty}N_{e}^{2}dr}
\nonumber \\
&=&\frac{\int_{R_{*}}^{\infty}N_{n}dr}
{\int_{R_{*}}^{\infty}N_{e}^{2}dr}\,,
\end{eqnarray}

\noindent based on the detailed NLTE analysis of a stellar wind
immersed in the radiation field of a central star, with statistical
balance between photoionization and radiative recombination.  The
departure coefficient for level n is then the ratio of the NLTE and
LTE values of $\overline{q}_{n}$ .

The calculation of $\overline{q}_{nLTE}$ is done as follows:

\begin{eqnarray}
\int_{R_{*}}^{\infty}N_{n}(r)dr&=&n^{2}N_{01}e^{X(n)-X(1)}\int_{R_{*}}
^{\infty}\left(\frac{R_{*}}{r}\right)^{2\eta} dr \nonumber \\
&=&\frac{n^{2}N_{01}R_{*}}{2\eta-1}e^{X(n)-X(1)}
\end{eqnarray}

\noindent and

\begin{eqnarray}
\int_{R_{*}}^{\infty}N_{e}^{2} dr&=&N_{0e}^{2}\int_{R_{*}}^{\infty}
\left(\frac{R_{*}}{r}\right)^{2\eta} dr \nonumber \\
&=&\frac{N_{0e}^{2}R_{*}}{2\eta-1}\,,
\end{eqnarray}

\noindent so that

\begin{eqnarray}
\overline{q}_{nLTE}&=&\frac{\int_{R_{*}}^{\infty}N_{n}dr}
{\int_{R_{*}}^{\infty}N_{e}^{2}dr} \nonumber \\
&=&\frac{N_{01}}{N_{0e}^{2}}n^{2}e^{X(n)-X(1)}
\nonumber \\
&=&(2\pi m_{e}kT_{e})^{-1.5}h^{3}n^{2}e^{X(n)}\,.
\end{eqnarray}

The population parameters $\overline{q}_{2NLTE}$ and
$\overline{q}_{3NLTE}$ may be estimated by extrapolation of power law
fits to the disk temperatures (10,000--20,000~K) and mass-loss
rates ($\sim10^{-8}~M_{\sun}~yr^{-1})$ characteristic of Be stars,
based on Figure~3 of CNM:

\begin{equation}
\overline{q}_{2NLTE}=1.11\times10^{-8}\,T_{e}^{-2.83}\,,
\end{equation}

\begin{equation}
\overline{q}_{3NLTE}=4.59\times10^{-13}\,T_{e}^{-2.02}\,.
\end{equation}

The approximate departure coefficients are then
$b_{2}=\overline{q}_{2NLTE}/\overline{q}_{2LTE}$ and
$b_{3}=\overline{q}_{3NLTE}/\overline{q}_{3LTE}$, so that the
corrected level populations are roughly
$N_{2NLTE}=b2 \times N_{2LTE}$ and $N_{3NLTE}=b3 \times N_{3LTE}$.
These correction factors are applied to the appropriate terms in the
summation inside the parentheses in equation~(5).  The corrected form
of $\kappa(\lambda)$ is used only in calculating the absorption
optical depth and not in calculating the emission coefficient, since
recombination and free-free emission are LTE processes.

\end{document}